\documentclass[12pt]{article}
\usepackage{epsfig}

\usepackage{amssymb}
\usepackage{amsmath}
\usepackage{amsfonts}
\usepackage{amsthm}

\theoremstyle{theorem}

\theoremstyle{definition}

\def\bp{\begin{proof}}
\def\ep{\end{proof}}

\def\be{\begin{equation}}
\def\ee{\end{equation}}
\def\ba{\begin{array}{c}}
\def\ea{\end{array}}

\newcommand{\bea}{\begin{eqnarray}}
\newcommand{\eea}{\end{eqnarray}}

\newcommand{\bbr}{\br\!\br}
\newcommand{\kkt}{\kt\!\kt}

\newcommand{\kt}{\rangle}
\newcommand{\br}{\langle}
\begin{document}

\titlepage

\vspace{.35cm}

 \begin{center}{\Large \bf

${\cal CPT}-$symmetric discrete square well.

  }\end{center}

\vspace{10mm}

 \begin{center}

 {\bf Miloslav Znojil}

 \vspace{3mm}
Nuclear Physics Institute ASCR,

250 68 \v{R}e\v{z}, Czech Republic

{e-mail: znojil@ujf.cas.cz}

\vspace{3mm}

and \vspace{3mm}

 {\bf Milo\v{s} Tater}

 \vspace{3mm}
Nuclear Physics Institute ASCR,

250 68 \v{R}e\v{z}, Czech Republic

{e-mail: tater@ujf.cas.cz}

\vspace{3mm}

%\today, rkiicr.tex

%of the form $V_1(x)=\lambda\,V_-(x) + \mu\,V_+(x)$

\end{center}

\vspace{5mm}

%\newpage

\section*{Abstract}

In an addendum to the recent systematic Hermitization of certain $N$
by $N$ matrix Hamiltonians $H^{(N)}(\lambda)$ \cite{all} we propose
an amendment $H^{(N)}(\lambda,\lambda)$ of the model. The gain is
threefold. Firstly, the updated model acquires a natural
mathematical meaning of Runge-Kutta approximant to a differential
${\cal PT}-$symmetric square well in which ${\cal P}$ is parity.
Secondly, the appeal of the model in physics is enhanced since the
related operator ${\cal C}$ of the so called ``charge" (the
requirement of observability of which defines the most popular
Bender's metric $\Theta={\cal PC}$) becomes also obtainable (and is
constructed here) in an elementary antidiagonal matrix form at all
$N$. Last but not least, the original phenomenological energy
spectrum is not changed so that the domain of its reality [i.e., the
interval of admissible couplings $\lambda\in (-1,1)$] remains the
same.

\newpage
 \section{Introduction }

In paper~\cite{all} (to be cited as paper I in what follows) we felt
inspired by the recently revealed formal merits of an exceptionally
easy non-trivial-metric-mediated Hermitizations of
boundary-condition interactions \cite{David}. We turned attention
there to the simplified, discrete, real and manifestly asymmetric
tridiagonal $N$ by $N$ matrix Hamiltonians
 \be
  H^{(N)}({\lambda,\mu})=  \left[ \begin {array}{cccccc}
 2&-1-{\it {\lambda}}&0&\ldots&0&0
\\
{}-1+{\it {\lambda}}&2&-1&0&\ldots&0
\\
{}0&-1&\ \ \ 2\ \ \ &\ddots&\ddots&\vdots
\\
{}\vdots&0&\ddots&\ \ \ \ddots\ \ \ &-1&0
\\
{}0&\vdots&\ddots&-1&2&- 1-{\it {\mu}}
\\
{}0&0&\ldots&0&-1+{\it {\mu}}&2
\end {array}
 \right]\,.
 \label{toymo}
 \ee
We studied these models under an additional, more or less randomly
selected simplification with $\mu \to -\lambda$ where we abbreviated
$H^{(N)}({\lambda,-\lambda})= H^{(N)}(\lambda)$.

In the present brief complement of paper I we intend to demonstrate
that in spite of certain inessential complications of mathematics,
the alternative simplifying choice of $\mu = +\lambda$ leads to an
enhanced appeal of the physics described by the alternative (and,
incidentally, isospectral) toy model $H^{(N)}({\lambda,\lambda})$.

After a very compact summary of the state of the art relocated to
Appendices A and B below, the detailed presentation of our new
results is split in section \ref{IIa} (explaining certain practical
aspects of the difference between the concepts of ${\cal
PT}-$symmetry and ${\cal P}-$pseudohermiticity), section \ref{IIb}
(clarifying the mathematical and physical meaning of the Bender's
\cite{Carl} special operator of ``charge" ${\cal C}$) and section
\ref{IIc} (where we show that for our updated Hamiltonians
$H^{(N)}({\lambda,\lambda})$ the ``charge" ${\cal C}$ proves
obtainable in an extremely elementary closed matrix form).

Section~\ref{sigmase} will summarize our results re-emphasizing that
our construction of ${\cal C}$ guarantees the physical acceptability
of the model. We shall point out that the availability of compact
formula for ${\cal C}$ renders our amended Hamiltonian
$H^{(N)}({\lambda,\lambda})$ fully compatible with all of the
postulates of standard quantum theory. In comparison with the model
of paper I, it is also better understood in the continuous limit of
$N \to \infty$.

\section{Parity ${\cal P}$\label{IIa}}

The initial encouragement to our present study stemmed from a few
observations as made in paper I. Without repeating the details here
we may merely recall the definition of the ${\cal PT}-$symmetry of
$H$ (cf. also Eq.~(\ref{pt}) in Appendix A below) and rewrite it in
its fully explicit equivalent matrix form
 \be
 \sum_{i=1}^N\,
 \left [
      \left (H^\dagger\right )_{ji}\,{\cal P}_{in}
      -{\cal P}_{ji}\,H_{in}\right ] =0
 \,,\ \ \ \ \ j,n=1,2,\ldots,N
   \,.
 \label{htotbe}
 \ee
In paper I we revealed that there exists the whole set of
sparse-matrix pseudometrics ${\cal P}={\cal P}_k^{(N)}$ with
$k=1,2,\ldots,N$, none of which appeared to be a rigorous discrete
approximant to the parity. At the same time, our ability of finding
{\em all} matrices ${cal P}$ compatible with $H$ via
Eq.~(\ref{htotbe}) facilitated significantly our discussion and
construction of the metrics.

The difficulty of the search for solutions ${\cal P}$ of
Eq.~(\ref{htotbe}) is one of the main obstacles of an exhaustive
understanding of {\em physics} which can {\em potentially} be
covered by {\em any} given Hamiltonian, Hermitian or not
\cite{Geyer}. In parallel, the search for the auxiliary,
sparse-matrix solutions ${\cal P}$ of Eq.~(\ref{htotbe}) (we may
call it Dieudonn\'e's equation \cite{Williams}) may be perceived as
one of the key {\em mathematical} conditions of practical
applicability of the vast majority of non-Hermitian $H \neq
H^\dagger$.

The particular minus-sign model $H^{(N)}(\lambda)
\,\equiv\,H^{(N)}({\lambda,-\lambda})$ of paper I admitted a {\em
complete} solution of Eq.~(\ref{htotbe}). From our present point of
view the serious drawback and weakness of the model lies in the fact
that {\em all} operators ${\cal P}$ compatible with the
Dieudonn\'e's equation proved {\em manifestly coupling-dependent}.

This is an unpleasant feature. For illustration let us recall
equation Nr.~(15) of paper I which defines the most natural (viz.,
antidiagonal) candidate for the parity-reminding operator compatible
with Eq.~(\ref{pt}). It possesses the indefinite-matrix closed form
 \be
 {\cal P}={\cal P}^{(N)}({\lambda})=
  \left[ \begin {array}{ccccc}
 0&0&\ldots&0&\alpha
 \\{}0&\ldots&0&1&0\\
 {}\vdots&
 {\large \bf _. } \cdot {\large \bf ^{^.}}&
 {\large \bf _. } \cdot {\large \bf ^{^.}}
 &
 {\large \bf _. } \cdot {\large \bf ^{^.}}&\vdots
 \\{}0&1&0&\ldots&0
 \\{}\alpha&0&\ldots&0&0
 \end {array} \right]\,,\ \ \ \ \alpha={\frac
 {1-{\it {\lambda}}}{1+{\it {\lambda}}}}\,
 \label{andiago}
 \ee
which varies with $\lambda$. In the continuous limit $N \to \infty$
(viz., Runge-Kutta limit, see paper I) this matrix cannot be
interpreted as a standard, coupling-independent operator of parity,
therefore. Strictly speaking, Hamiltonian
$H^{(N)}({\lambda,-\lambda})$ remains out of the scope of ${\cal
PT}-$symmetric quantum mechanics. This observation motivated our
present study.

In a search for a manifestly ${\cal PT}-$symmetric model with better
properties we were rather lucky when we turned attention to the next
simplest choice of the parameters and inserted $\mu=+\lambda$ in
Eq.~(\ref{toymo}). The core of the success lied in the emergence and
verification of the coupling-independence of the most natural
candidate
 \be
 {\cal P}={\cal P}^{(N)}(0)=
  \left[ \begin {array}{ccccc}
 0&0&\ldots&0&1
 \\{}0&\ldots&0&1&0\\
 {}\vdots&
 {\large \bf _. } \cdot {\large \bf ^{^.}}&
 {\large \bf _. } \cdot {\large \bf ^{^.}}
 &
 {\large \bf _. } \cdot {\large \bf ^{^.}}&\vdots
 \\{}0&1&0&\ldots&0
 \\{}1&0&\ldots&0&0
 \end {array} \right]\,
 \label{anago}
 \ee
for the pseudometric operator entering Eq.~(\ref{htotbe}) and/or
Eq.~(\ref{pt}) below.

\section{The operator of charge \label{IIb}}

A hidden root of success of ${\cal PT}-$symmetric models in physics
may be seen in the decisive simplification of the theory using a
specific form of the third, ``superfluous" representation space
${\cal H}^{(S)}={\cal H}^{(S)}_{(\cal CPT)}$. The metric
$\Theta=\Theta_{(\cal CPT)}$ is only considered in a specific
factorized form $\Theta_{(\cal CPT)}\,\equiv\,{\cal CP}$. Under such
a simplifying assumption one reveals the equivalence of
Eq.~(\ref{ptc}) of Appendix A and the formula representing the
Hermiticity of the model in space ${\cal H}^{(S)}$. In its matrix
form this equation reads
 \be
 \sum_{k=1}^N\,
 \left [
      \left (H^\dagger\right )_{jk}\,\Theta_{kn}
      -\Theta_{jk}\,H_{kn}\right ] =0
 \,,\ \ \ \ \ j,n=1,2,\ldots,N
   \,,
 \label{htot}
 \ee
is linear in $\Theta$ and very similar to Eq.~(\ref{htotbe}) of
preceding section. Thus, one may make use of the $N-$parametric
ansatz
 \be
   \Theta^{(N)}=
   \sum_{k=1}^N\, {\mu_k}\,{\cal P}_k^{(N)}\,
 \label{e777}
   \ee
where just the positivity of $\Theta^{(N)}$ must be required and
guaranteed (cf. \cite{fund}).

In the above-outlined context the reconstruction of the charges
might proceed via the knowledge of the sums
 \be
 \Theta=\sum_k\,|k\kkt\, \kappa_k^2\,\bbr k|\,,\ \ \ \ \ \
 {\cal P}^{-1} = \sum_m\,|m\kt\, \nu_m\,\br m|\,
 \ee
where the single-ketted and double-ketted symbols $|k\kt$ and
$|k\kkt$ denote the $k-$th eigenvectors of $H$ and $H^\dagger$,
respectively. This strategy (employed, say, in Refs.
\cite{David,David1}) will be used also in what follows. In more
detail, once the real parameters $\kappa_k$ and $\nu_m$ stay
variable and virtually arbitrary (cf. \cite{SIGMAbe} for details)
one can postulate, in our biorthogonal basis, the spectral formula
 \be
 {\cal C} = \sum_n\,|n\kt\, \omega_n\,\bbr n|\,.
 \ee
The (in general, complex) values of overlaps $\mu_n=\bbr n|n\kt$ may
be considered known. Constraints ${\cal C}={\cal P}^{-1}\Theta$ and
${\cal C}^2=I$ then merely imply that we have to demand that
 \be
 \omega=\mu_n^*\nu_n\kappa_n^2\,,\ \ \ \ \ \
 \mu_n\,\omega^2_n=1/\mu_n\,.
 \label{jedub}
 \ee
For a given set of ``input" data $\mu_n$ and $\nu_n$ this determines
all the coefficients $\kappa_n^2>0$ so that only a change in our
choice of parity ${\cal P}$ may lead to a different version of
${\cal C}$ and of the metric.

In this sense, the undeniable phenomenological as well as
theoretical appeal of ${\cal PT}-$symmetric models lies in the
existence of a straightforward recipe for suppression of the well
known ambiguity of the assignment of the metric $\Theta$ to a given
${\cal PT}-$symmetric Hamiltonian. In matrix models the operator
${\cal C}$ is not a sparse matrix while it {\em proved to be} a
sparse in our present, exceptional model $H^{(N)}(\lambda,\lambda)$.

\section{Charge ${\cal C}$ for
Hamiltonian $H^{(N)}({\lambda,\lambda})$\label{IIc}}

The privileged metrics of the form $\Theta={\cal PC}$ are usually
accepted on the purely pragmatic grounds of simplicity. In general,
one still must get through complicated calculations before arriving
at any concrete ${\cal C}$ or $\Theta={\cal PC}$. An exception has
been found in paper I where a systematic construction of {\em all}
the eligible $\Theta$s has been shown feasible due to a maximal
friendliness of the model $H=H^{(N)}({\lambda,-\lambda})$.

Beyond the framework of  ${\cal PT}-$symmetric quantum mechanics
which requires the observability of charge the formal assignment of
the metric to a Hamiltonian requires an alternative specification of
the menu of required observables \cite{Geyer,conceptual}. In this
sense, the additional requirement of the existence of charge ${\cal
C}$ may be perceived as one of the most compact recipes for making
the model unambiguous.

In our present letter we accepted such a research project, made use
of the symbolic-manipulation algorithms developed in paper I and
applied them to the alternative model
$H=H^{(N)}({\lambda,+\lambda})$. As long as the updated equations
(\ref{htotbe}) exhibit now much less symmetries, we were really
surprised by revealing that our rather naive strategy gave us an
affirmative answer of an unexpectedly elementary form. The resulting
charge has been found in the following, purely antidiagonal and
manifestly involutive form
 \be
 {\cal C}^{(N)}({\lambda})=
  \left[ \begin {array}{ccccc}
 0&0&\ldots&0&1/\alpha({\lambda})
 \\{}0&\ldots&0&1&0\\
 {}\vdots&
 {\large \bf _. } \cdot {\large \bf ^{^.}}&
 {\large \bf _. } \cdot {\large \bf ^{^.}}
 &
 {\large \bf _. } \cdot {\large \bf ^{^.}}&\vdots
 \\{}0&1&0&\ldots&0
 \\{}\alpha({\lambda})&0&\ldots&0&0
 \end {array} \right]\,,\ \ \ \ \alpha(\lambda)={\frac
 {1-{\it {\lambda}}}{1+{\it {\lambda}}}}\,.
 \label{aiago}
 \ee
This corresponds to the safely positive and purely diagonal metric
 \be
 \Theta^{(N)}_{0}({\lambda})=
  \left[ \begin {array}{ccccc}
 \alpha({\lambda}) &0&\ldots&0&0
 \\{}0&1&0&\ldots&0\\
 {}\vdots&\ddots&\ddots&\ddots&\vdots
 \\{}0&\ldots&0&1&0
 \\{}0&0&\ldots&0&1/\alpha({\lambda})
 \end {array} \right]\,,\ \ \ \ \alpha(\lambda)={\frac
 {1-{\it {\lambda}}}{1+{\it {\lambda}}}}\,.
 \label{diso}
 \ee
The latter metric is only unessentially more complicated than its
maximally symmetric diagonal analogue as specified by equation
Nr.~(14) in paper~I.

%\newpage
 \section{Summary \label{sigmase}}

In a brief conclusion let us emphasize that the exact solvability of
models exhibiting hidden Hermiticity must involve {\em not only} the
feasibility of diagonalization of the Hamiltonian {\em but also} the
feasibility of construction and selection of an optimal metric
$\Theta$. There exist not too many quantum models which would
satisfy both these criteria. Among them we were inspired  by the
models with $N=\infty$ in which the unitarity of the quantum
scattering has been achieved~\cite{scatt}.

We paid attention just to bound states here. Naturally, our work has
significantly been simplified by the results of paper I because due
to the easily demonstrated isospectrality relationship between
Hamiltonians $H^{(N)}(\lambda)=H^{(N)}({\lambda,-\lambda})$ and
$H^{(N)}({\lambda,+\lambda})$ the spectrum of energies
$E^{(N)}_n({\lambda})$ remains real and non-degenerate for the same
set of couplings lying inside the same open and $N-$independent
interval of $\lambda \in  (-1,1)$.

In the updated calculations devoted to $H^{(N)}({\lambda,+\lambda})$
we succeeded in satisfying {\em both} the constraints of ${\cal
PT}-$symmetry and ${\cal CPT}-$symmetry {\em exactly}, in
non-perturbative manner. We may summarize that

\begin{itemize}

\item
our Hamiltonians $H^{(N)}({\lambda,\lambda})$ are ${\cal
PT}-$symmetric in the narrow sense, satisfying relation (\ref{pt})
of Appendix A below with $\lambda-$independent parity~(\ref{anago});

\item
for the same model there  exists the operator of charge ${\cal C}$
such that ${\cal C}^2=I$ which is represented by an extremely
elementary matrix~(\ref{aiago}).

\end{itemize}

 \noindent
In the spirit of review \cite{Carl} and
Refs.~\cite{conceptual,BBJ,Erratum}, our explicit construction of
charge makes the corresponding amended toy model manifestly ${\cal
CPT}-$symmetric, i.e., compatible with the constraint (\ref{ptc}) of
Appendix A below. This means that every Hamiltonian
$H^{(N)}({\lambda,\lambda})$ with the not too strong coupling
$\lambda \in (-1,1)$ is assigned an exceptional physical Hilbert
space ${\cal H}^{(S)}_{({\cal CPT})}$ in which the metric is defined
as  product $\Theta^{(N)}_{({\cal CPT})}={\cal PC}$.

\section*{Acknowledgement}

Work supported by the M\v{S}MT ``Doppler Institute" project Nr.
LC06002 and by the Institutional Research Plan AV0Z10480505.

% \newpage

\newpage

\section*{Appendix A: The concepts of ${\cal PT}$ and ${\cal CPT}$ symmetries }

During the recent developments of quantum mechanics of (stable)
bound states as initiated by Bender and Boettcher \cite{BB} it
became increasingly popular to select a suitable phenomenological
potential $V(x)$ in an innovative non-selfadjoint version. In such a
purely theoretical setting (as reviewed, e.g., in Refs.~\cite{Carl})
as well as in its very recent experimental verifications
\cite{Kottos} it makes sense to choose $V(x)$ in an exactly solvable
and simplest possible form, say, of a square well \cite{sqw} or of a
point interaction \cite{Kurasov} or of their
perturbations~\cite{Tretter}.

In all of the similar non-selfadjoint models it is generically
difficult to complement the ``easy" results concerning energies by
the ``difficult" predictions of the role and physical interpretation
of any other observable quantity (like, e.g., of the coordinate
\cite{Batal}). In the context of nuclear physics, for example, this
problem of interpretation has already been addressed almost twenty
years ago \cite{Geyer} but it still waits for a final satisfactory
resolution \cite{SIGMA,fund}.

One of the most widely accepted transient strategies lies in the use
of constructions of a single additional observable ${\cal C}$ with
eigenvalues one or minus one. In accord with the original proposal
by authors of Ref.~\cite{BBJ} we may call it a charge. In general,
even such a simplified strategy need not imply its easy
implementation (cf., e.g., Ref.~\cite{cubic} for an illustration).
One has to feel satisfied by any partial success.

A sample of such a success has been described in Ref.~\cite{David}
where the operator ${\cal C}$ was constructed as a product of parity
${\cal P}$ and of the so called metric $\Theta$. The feasibility of
the construction (based on a resummation of infinite series)
required a maximal simplicity of the interaction. Thus, the deep
one-dimensional square-well potential $V_0(x)$ was assumed perturbed
just by a user-friendly point interaction $V_1(x)=\lambda\,V_-(x) +
\mu\,V_+(x)$ acting at the two boundaries of the well. Even after
such a comparatively drastic simplification (after which the
interaction proved equivalent to a mere redefinition of boundary
conditions) the analysis of the model still remained quite difficult
(cf. also its amended version in~\cite{David1}).

A technically less difficult approach to the construction of ${\cal
C}$ has been found after a further transition from the differential
Hamiltonian operators to their discrete approximants and analogues
\cite{disqw}. A new broad and interesting class of tractable quantum
models emerged. Even the extremely schematic $N-$dimensional matrix
models with $N=2$ proved of value for our understanding of certain
theoretical subtleties \cite{plb,bila,mann}.

Once one restricts attention to Jacobi-matrix models exemplified by
Eq.~(\ref{toymo}) and possessing a nontrivial interaction terms
solely near the ends of the main diagonal, the first important merit
of the resulting matrix toy Hamiltonian appeares to be the reality
of the energy spectrum inside a non-empty domain ${\cal D}$ of
parameters $\lambda$ and $\mu$. In Figures 1 - 6 of paper~I a few
characteristic samples of the rich and flexible parametric
dependence of this spectrum were displayed. Under additional
restriction $\mu=\pm \lambda\ $  domains ${\cal D}$ were shown
$N-$independent and specified as a sufficiently large interval of
$\lambda \in (-1,1)$. Within the latter interval we proceeded in a
constructive manner, selected a definite (viz., minus) sign in
$H=H^{(N)}({\lambda,- \lambda})$ and constructed an {\em exhaustive}
list of $N-$dimensional, not necessarily positive definite
pseudometrics ${\cal P}={\cal P}^\dagger$. In the spirit of review
papers \cite{Alirep} these operators were interpreted as alternative
generalized parities.

In the light of an older mathematical study \cite{Dieudonne} the
latter operators ${\cal P}$ need not even be required invertible.
Nevertheless, all of them were designed as compatible with the
${\cal PT}-$symmetry constraint imposed upon the Hamiltonian and
written in the form
 \be
 {\cal PT}\,H^{(N)}({\lambda,\mu})=
 H^{(N)}({\lambda,\mu})\,{\cal PT}\,
 \label{pt}
 \ee
Although operator ${\cal T}$ is intended to simulate time reversal,
a vivid debate in the literature \cite{Erratum} indicated that some
of its technical aspects are nontrivial. The net result of this
debate may be summarized as a conclusion that the existence of
symmetry (\ref{pt}) facilitates a safe return to the traditional
textbook formalism of quantum theory.

Let us add here also the well known fact that the non-Hermiticity of
Hamiltonians with real spectra may be treated as just a
misinterpretation or rather a price paid for a wrong choice of the
Hilbert space. In one of the most popular resolutions of the
apparent paradox Carl Bender with coauthors \cite{Carl} introduced
the concept of charge ${\cal C}$ with the only (i.e., multiply
degenerate) eigenvalues equal to $+1$ or $-1$. The availability of
this charge enabled them to introduce an amended, standard Hilbert
space where the input Hamiltonian $H$ becomes self-adjoint. The
recipe has been shown equivalent to the additional symmetry
requirement
 \be
 {\cal CPT}\,H^{(N)}({\lambda,\mu})=
 H^{(N)}({\lambda,\mu})\,{\cal CPT}\,.
 \label{ptc}
 \ee
The important role of this constraint contrasts with the scarcity of
the available known pairs of mutually compatible observables $H$ and
${\cal C}$. In this sense our present paper partially fills the gap.

\section*{Appendix B. Hidden Hermiticity }

The abstract formalism of quantum theory is frequently being
explained via concrete descriptions of a point particle moving in a
confining one-dimensional potential well $V(x)$ \cite{Fluegge}. In
the notation of our review \cite{SIGMA} one prefers the use of a
specific, ``friendly" realization ${\cal H}^{(F)}$ of the abstract
Hilbert space of states. In it, the measurable coordinate $q \in
\mathbb{R}$ (i.e., strictly speaking, the eigenvalue of operator
$Q^{(F)}$ of observable position) {\em coincides} with the argument
$x$ of the normalized bound-state wave function
$\psi_n=\psi_n^{(F)}(x)$ (cf. also a longer exposition of this
subtlety in \cite{Hoo}).  In the Dirac's compact notation one
prefers working with the ket vectors $ |\psi_n^{(F)}\kt \in {\cal
H}^{(F)}\ \equiv \ \mathbb{L}^2(\mathbb{R})$.

For non-local potentials $V(x,x')$ one may turn attention to an
alternative, momentum representation $|\psi_n^{(P)}\kt \in {\cal
H}^{(P)}\ \equiv \ \mathbb{L}^2(\mathbb{R})$ of the same states. The
Fourier-like mapping ${\cal F}$ between spaces ${\cal H}^{(F)}$ and
${\cal H}^{(P)}$ is postulated unitary, ${\cal F}^\dagger {\cal F} =
I$. The physical meaning of the real line $\mathbb{R}$  is now
different but the physical contents of the theory remains the same.

In a generalization of this picture one replaces operators ${\cal
F}$ by non-unitary maps $\Omega$ and defines
 \be
 |\psi_n^{(P)}\kt = \Omega\,|\psi^{(F)}\kt
 \,,
 \ \ \ \ \ \Omega^\dagger \Omega = \Theta \neq I\,.
 \label{trik}
 \ee
Taken, originally, as a mere mathematical curiosity
~\cite{Dieudonne,BG} the latter trick  proved unexpectedly useful
and fruitful in nuclear physics where the first non-unitary version
of the boson-fermion map $\Omega$ has been proposed by Dyson (cf.
review \cite{Geyer}). Its mathematical essence may most easily be
interpreted as an introduction of certain third representation space
${\cal H}^{(S)}$ where superscript $^{(S)}$ may stand for
``standard" or, if you wish, ``sophisticated".

By construction, the two Hilbert spaces ${\cal H}^{(S)}$ and ${\cal
H}^{(P)}$ must be unitarily equivalent. Similar requirement does not
apply to the two spaces ${\cal H}^{(S)}$ and ${\cal H}^{(F)}$ which
may only be allowed to coincide as Banach spaces formed by the
identical vector spaces ${\cal V}^{(F)}={\cal V}^{(S)}$ of kets,
with their inner products not yet specified. In the latter pair the
differences only emerge between the dual (i.e.,
representation-dependent and, in standard notation, primed) Banach
spaces of bra-vectors. The first space $\left ({\cal V}^{(F)}\right
)'$ of linear functionals in ${\cal H}^{(F)}$ must be different from
the second space $\left ({\cal V}^{(S)}\right )'$ of linear
functionals in ${\cal H}^{(S)}$.

%\subsection{ \label{IV}}

In order to avoid confusion we would recommend the use of the
full-fledged notation of Ref. \cite{Gegen}. In it, the respective
operations of Hermitian conjugation~${\cal T}={\cal T}^{(F,P,S)}$
are defined as the most common vector (or matrix) transposition plus
complex conjugation  in ${\cal H}^{(F)}$,
 \be
 {\cal T}^{(F)}: |\psi_n^{(F)}\kt \ \to  \ \br \psi_n^{(F)}|\ \in
 \  \left ({\cal V}^{(F)}\right )'\,\,
 \label{bubu}
 \ee
and in ${\cal H}^{(P)}$,
 \be
 {\cal T}^{(P)}:  |\psi_n^{(P)}\kt \ \to  \ \br \psi_n^{(P)}|\ \in
 \  \left ({\cal V}^{(P)}\right )'\,
 \ee
and as the more complicated prescription valid in ${\cal H}^{(S)}$,
 \be
 {\cal T}^{(S)}: |\psi_n^{(S)}\kt \ \to  \ \br \psi_n^{(S)}|
 = \br \psi_n^{(F)}|\,\Theta \ \equiv \
 \bbr \psi_n| \ \in \
  \left ({\cal V}^{(S)}\right )'\,.
  \label{zebu}
 \ee
A more extensive discussion of such a version of quantum theory in
its triple Hilbert-space representation may be found in
Ref.~\cite{SIGMA}.

\end{document}